\input{epsf}

\documentstyle{mn}

\newcommand{\kms}{\mbox{km}\,\mbox{s}^{-1}}
\newcommand{\sbin}{\mbox{km}^2\,\mbox{s}^{-2}}
\newcommand{\skm}{\mbox{s}\,\mbox{km}^{-1}}

\newcommand{\vtran}{v_{\rm t}}
\newcommand{\nsmall}{n_{\vtran}}
\newcommand{\nbig}{N_{\vtran}}

\title[Velocity distribution of stars in the solar neighbourhood]
   {Velocity distribution of stars in the solar neighbourhood
   \thanks{Based on data from the ESA Hipparcos astrometry satellite}}

\author[J. Skuljan, J.B. Hearnshaw and P.L. Cottrell]
       {J. Skuljan, J.B. Hearnshaw and P.L. Cottrell \\
        Department of Physics and Astronomy, University of Canterbury,
           Private Bag 4800, Christchurch, New Zealand}
\date{Accepted 1999 April 10.
      Received 1999 April 10;
      in original form 1998 October 12}

\pagerange{\pageref{firstpage}--\pageref{lastpage}}

\pubyear{1999}

\begin{document}

\maketitle

\label{firstpage}

\begin{abstract}
A two-dimensional velocity distribution 
in the $UV$-plane has been obtained
for stars in
the solar neighbourhood, using the Hipparcos astrometry for
over 4000 `survey' 
stars with parallaxes greater than 10 mas and radial velocities found in 
the Hipparcos Input Catalogue.
In addition to the already known grouping characteristics (field 
stars plus young
moving groups), the velocity distribution seems to exhibit a more complex
structure characterized by several longer {\it branches} running almost
parallel to each other across the $UV$-plane. By using the wavelet
transform technique to analyse the distribution, 
the branches are visible at relatively
high significance levels of 90 per cent or higher. They are
roughly equidistant with a separation of about 15~$\kms$ for early-type
stars
and about 20~$\kms$ for late-type stars, creating
an overall quasi-periodic structure which can
also be detected by means of a two-dimensional Fourier transform. 
This branch-like velocity distribution 
might be due to the galactic spiral structure.
\end{abstract}

\begin{keywords}
Galaxy: kinematics and dynamics -- solar neighbourhood.
\end{keywords}

\section{Introduction}

The velocity distribution of stars in the solar neighbourhood is an
important tool for studying different aspects of galactic kinematics and
dynamics. During a long era of ground-based astrometry that preceded
the Hipparcos mission, many subtle details in the velocity field have
gone undetected due to the smearing caused by large
uncertainties in stellar parallaxes. 
With the Hipparcos measurements
we are in a 
position to investigate the structure in more
detail, 
confirming some previous characteristics and discovering some new features.

A typical analysis of the distribution of stars in the $UV$-plane
concentrates on determining the {\it velocity ellipsoid} and its
parameters (dispersions in $U$, $V$ and $W$, as well as the orientation
of the principal axis, i.e.~the longitude of the vertex). More
details about this can be found in any textbook on galactic structure
and
kinematics
(e.g.~Mihalas \& Binney 1981). In addition to the
global properties of the velocity ellipsoid, a variety of different
local irregularities are also studied. Certain concentrations of stars
in velocity space mean that there exist groups of stars ({\it moving
  groups}) that move with the same
velocity. This idea has been elaborated extensively in works by 
Eggen (e.g.~1970) and other authors.
Different authors use different techniques and different
stellar samples, but they all report the presence of moving groups in
the solar neighbourhood (Figueras et al.~1997,
Chereul et al.~1997,1998,1999, Dehnen 1998, Asiain et al.~1999).

This paper is part of a larger project started at the University of
Canterbury in order to test Eggen's hypothesis 
(Skuljan et al.~1997).
Here we discuss some inhomogeneities in the velocity
distribution that are related to moving groups and can
give some clues to the problem of star formation.

\section{The sample}

For this study a sample of 4597 stars has been constructed using the 
following selection criteria:
\begin{enumerate}
   \item Parallax greater than 10 mas (stars within 100~pc
     of the Sun), and $\sigma_\pi/\pi<0.1$, as taken from the Hipparcos
     Catalogue (ESA 1997). 
   \item Survey flag (Hipparcos field H68) set to `S'.
   \item Existing radial velocities in the Hipparcos Input Catalogue
     (ESA 1992).
   \item Existing $B-V$ colours in the Hipparcos Catalogue.
\end{enumerate}

The survey flag has been used 
so that no stars 
proposed by
various individual projects are included in this analysis,
since they could introduce a bias.
Stars with parallax uncertainties greater than 10 per
cent have been rejected in order to have a reliable error propagation.
The Hipparcos $B-V$ colour
index is used here as a suitable indicator for dividing the sample into
two subsets, as explained in
Section~\ref{analysis}.

There are 12520 `survey' stars with parallaxes greater than 10 mas, out of 
118218 entries found in the 
Hipparcos Catalogue. However, only 11009 of these stars have their
parallax uncertainties less than 10 per cent. Finally, for 11007 of them
the $B-V$ colours are known.
On the other hand, we find 19467 stars with known
radial velocities, out of 118209 entries in the Hipparcos Input
Catalogue. Only 4597 stars are found in both subsets, if the
catalogue running numbers are used as a matching criterion 
($\mbox{HIP}=\mbox{HIC}$).

\begin{figure}
   \epsfxsize=\hsize\epsfbox{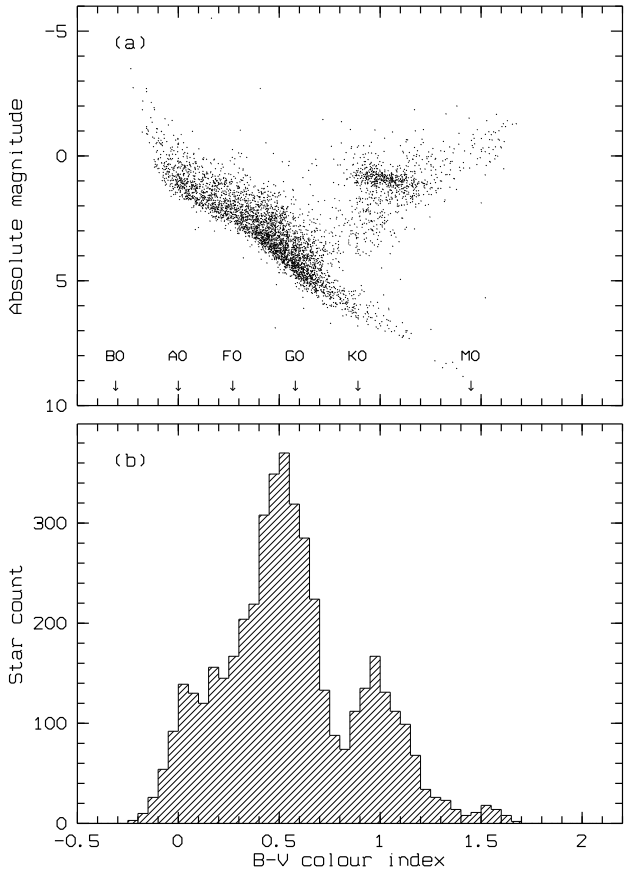}
   \caption{H-R diagram of the 4597 Hipparcos `survey' stars within
     100~pc, with known
     radial velocities and colours (a) and the distribution of the $B-V$
     colour index (b). The spectral classes (a)
     correspond to the main sequence (Allen 1991).}
   \label{hrdata}
\end{figure}

Before we proceed with our analysis of the velocity distribution, some
important points have to be emphasized here concerning the problem of
bias. First of all, not all spectral classes are equally represented in
our sample. The Hipparcos catalogue is essentially magnitude
limited, which means that we shall have a significant deficiency of red
dwarfs, compared to the young early-type stars. The situation is
illustrated in Figure~\ref{hrdata}. 
A great majority of stars are concentrated around $B-V=0.5$, corresponding 
to the main-sequence F stars. There is also a possible concentration of
earlier-type stars around A0, as well as a distinct peak of K
giants (red-clump stars on the horizontal branch, to be more precise)
around $B-V=1.0$. This should be kept in mind when drawing any
conclusions regarding the stellar ages (see Section~\ref{analysis}), but 
it will essentially not affect our results.

A possibly
more serious problem concerning our stellar sample is a {\it
  kinematic bias}. Binney et al.~(1997) demonstrated that radial
velocities are predominantly known for high-proper-motion stars. 
If only the stars with known radial velocities are used, then any
velocity distribution derived from such a biased sample might give a
false picture and lead to some wrong conclusions about the local stellar 
kinematics. That is the reason why many authors today choose not to include the
measured radial velocities at all (see also Dehnen \& Binney 1998,
Dehnen 1998, Cr{\'e}z{\'e} et al.~1998, Chereul et al.~1998,1999).

\begin{figure}
   \epsfxsize=\hsize\epsfbox{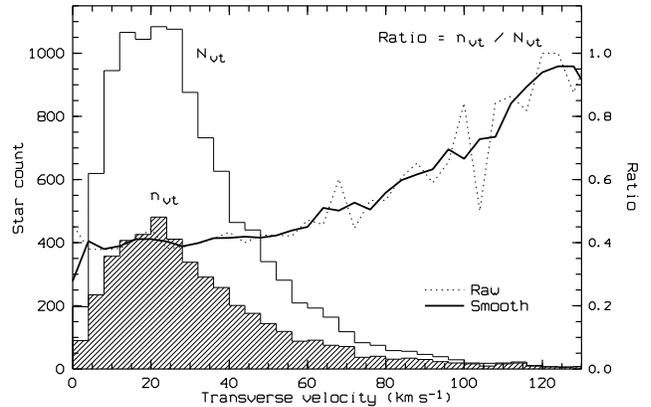}
   \caption{Kinematic bias, presented as a ratio between two
     distributions: stars with known radial velocities ($\nsmall$)
     out of the
     total sample of Hipparcos `survey' stars within 100~pc ($\nbig$),
     where $\vtran$ is the transverse velocity. Radial velocities are
     predominantly known for high-$\vtran$ stars, but the effect is
     important only at $\vtran > 70-80\ \kms$.}
   \label{rvbias}
\end{figure}

We have checked for potential kinematic bias in our case, and the result 
is presented in Figure~\ref{rvbias}. Two distributions are shown as
functions of the transverse velocity ($\vtran=4.74\,\mu/\pi$), one
for the total sample of 11007 `survey' stars within 100~pc ($\nbig$), 
and the other for the stars with known radial velocities only 
($\nsmall$). The $\vtran$-axis has been divided into 4-$\kms$
bins and the stars have been counted in each bin. If there was no
kinematic bias, then the probability that a star has a radial velocity
should be constant from bin to bin, and the ratio 
$\nsmall/\nbig$ would appear as
a flat line. It is obvious from Figure~\ref{rvbias} that this is not the 
case for our sample. 
While the radial velocities are known for about 40 per cent of the 
stars at $\vtran=20\ \kms$, the ratio reaches 80 per cent at 
$\vtran=120\ \kms$. However, the effect becomes a real problem only
at {\it higher velocities}, above 70 -- 80 $\kms$, as easily seen from
the diagram.
Below this limit, the 
ratio stays more or less flat, so that we can expect no significant
distortions in the inner parts of the velocity distribution, where we
shall primarily concentrate our attention. We shall return to this
problem again when we compare our velocity distribution to 
those of other authors 
(see Section~\ref{distribution}).

\section{Velocity distribution}
\label{distribution}

\begin{figure}
   \epsfxsize=\hsize\epsfbox{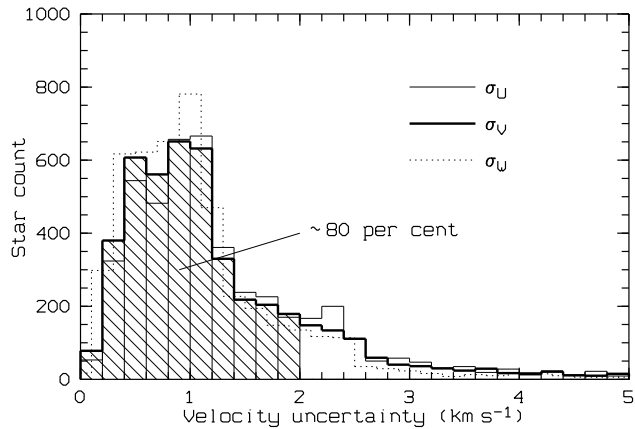}
   \caption{The distribution of the uncertainties in all three 
           velocity components: $\sigma_U$ (thin line),
            $\sigma_V$ (bold line) and $\sigma_W$ (dotted line).
           The uncertainty is less than $2\ \kms$ for about 80 per cent
           of stars.}
   \label{uvwsig}
\end{figure}

Having fixed the stellar sample, we can now proceed with the analysis.
Hipparcos parallaxes and proper motions, 
together with the radial velocities from the Hipparcos Input Catalogue, 
have been used to compute the stellar
space velocities relative to the Sun. 
The right-handed coordinate system has been used, with
the $X$-axis pointing towards the galactic centre, $Y$-axis in the
direction of galactic rotation (clockwise, when seen from the north
galactic pole), and $Z$-axis towards the north galactic
pole. Corresponding velocity components are $U$, $V$ and $W$
respectively. A typical error-bar in each velocity component is close to
$1\ \kms$, with about 80 per cent of stars having their velocity uncertainties
less than $2\ \kms$, as shown in Fig.~\ref{uvwsig}. The error propagation
has been treated taking into account the full correlation 
matrix between the Hipparcos astrometric 
parameters.

In order to estimate the probability density function $f(U,V)$ from the
observed data, we use here an {\it adaptive kernel method} (for more
details see Silverman 1986). The basic idea of this method is to apply a
smooth weight function (called the {\it kernel} function) to estimate the
probability density at any given point, using a number of neighbouring
data points. The term `adaptive' here means that the kernel width
depends on the actual density, so that the smoothing is done over a
larger area if the density is smaller.

\begin{figure}
   \epsfxsize=\hsize\epsfbox{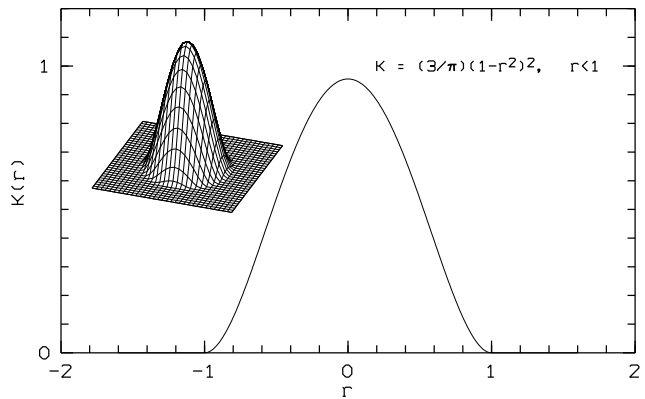}
   \caption{Kernel function $K(r)$ used for the 
            probability density estimation.}
   \label{kernel}
\end{figure}

We use the following definition of the adaptive kernel estimator (see
page 101 of Silverman 1986), defined 
at an arbitrary point $\vec\xi=(U,V)$:
\[
   \hat f(\vec\xi) = \frac{1}{n}\sum\limits_{i=1}^n
      \frac{1}{h^2\lambda_i^2}
         K\left(\frac{\vec\xi-\vec\xi_i}{h\lambda_i}\right)
\]
where $K(\vec r)$ is the kernel function, $\lambda_i$ are the local
bandwidth factors (dimensionless numbers controlling the overall kernel
width at each data point), and $h$ is a general smoothing factor.
We assume also that there are $n$ data points $\vec\xi_i=(U_i,V_i)$. Our 
function $K(\vec r)$ is a 2-D radially symmetric version of the {\it
  biweight} kernel (Fig.~\ref{kernel}), and is defined by:
\[
   K(r) = \left\{
      \begin{array}{ll}
         \frac{3}{\pi}(1-r^2)^2, & r<1 \\
         0, & r\geq1
      \end{array}
          \right.
\]
so that $\int K(\vec r)d\vec r=1$ (a condition that any kernel must
satisfy in order to produce an estimate $\hat f$ as a proper probability 
density function).

The local bandwidth factors $\lambda_i$ needed for the computation of
$\hat f(\vec\xi)$ are defined by:
\[
  \lambda_i = \left[\frac{\hat f(\vec\xi_i)}{g}\right]^{-\alpha}
\]
where $g$ is the geometric mean of the $\hat f(\vec\xi_i)$:
\[
\ln g = \frac{1}{n}\sum\limits_{i=1}^n\ln\hat f(\vec\xi_i)
\]
and $\alpha$ is a {\it sensitivity
parameter}, which we fix at $\alpha=0.5$ (a typical value for the
two-dimensional case). Note that in order to compute $\lambda_i$ we need 
the distribution estimate $\hat f$ which, in turn, can be computed only
when all $\lambda_i$ are known.
This problem, however, can be solved
iteratively, by starting with an approximate distribution (a fixed
kernel estimate, for example), and then improving the function as well
as the $\lambda_i$ factors in a couple of subsequent iterations.

Finally, an optimal value for the smoothing parameter $h$ is determined
using the least-squares cross-validation method, by minimizing the score 
function:
\[
   M_o(h)=\int\!\!\hat f^2\ -\ \frac{2}{n}
             \sum\limits_{i=1}^n\hat f_{-i}(\vec\xi_i)
\]
where $\int\!\!\hat f^2$ can be computed numerically, and
$\hat f_{-i}(\vec\xi_i)$ is the density estimate  at $\vec\xi_i$,
constructed from all data points {\it except} $\vec\xi_i$. It can be
shown that minimizing $M_o(h)$ is {\it equivalent} 
(in terms of mathematical expectation) 
to minimizing the integrated square error $\int(\hat f-f)^2$, so that
our estimate $\hat f$, based on the optimal value for $h$, is
as close as possible
to the true distribution $f$, using the data set available. In our case 
(for the whole sample of stars),
we have found the minimum of $M_o(h)$ at $h=10.7\ \kms$.

\begin{figure}
   \epsfxsize=\hsize\epsfbox{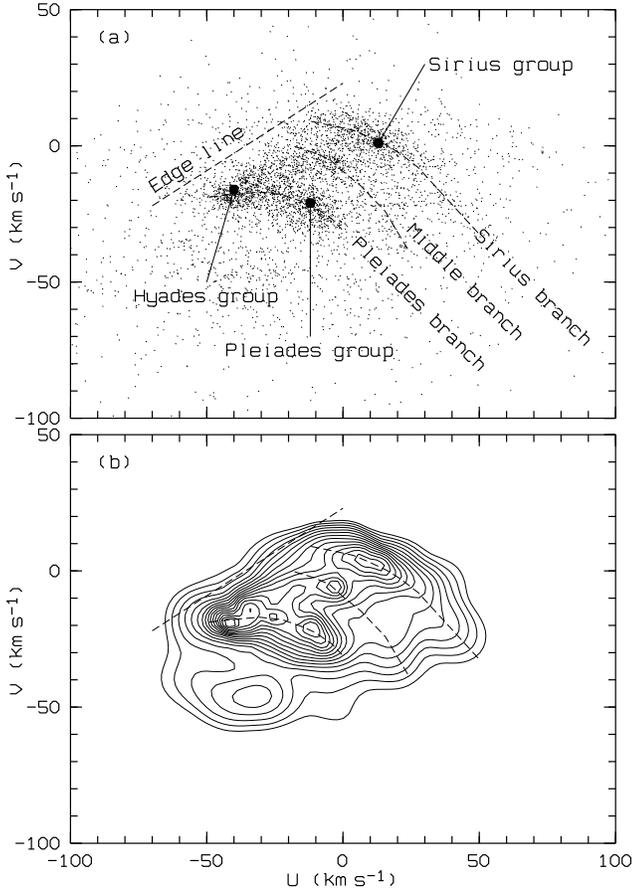}
   \caption{The distribution of stellar velocities in the $UV$-plane,
     shown as a scatter plot (a), and a contour
     plot of the probability density function (b) in logarithmic scale.
     The density function
     has been obtained using the adaptive kernel method (see text).
     The three main moving groups (a)
     are marked as filled circles. The three
     hypothetical branches and the 'edge line' 
     are shown as dashed lines.}
   \label{uvdata}
\end{figure}

Our $UV$-distribution of stellar velocities with respect to the Sun 
is shown in Fig.~\ref{uvdata}, both as a scatter plot and a smooth
contour plot representing the density function $\hat f(U,V)$, as computed using
the adaptive kernel method described above. The computations have been
performed on a grid of square bins of $2\times2$~$\kms$. This choice
for the bin size has been made taking into account a typical uncertainty 
in the velocity components, as mentioned earlier
(Fig~\ref{uvwsig}). Finally, the density function has been rescaled by a 
multiplication factor $nS$, where $n=4597$ is the total number of stars and
$S=4$~$\sbin$ is the area covered by a square bin. The numerical value
of the transformed distribution at any given bin is therefore equal to
the average number of stars falling in that bin (assuming that our
estimate $\hat f$ is close to the real distribution $f$).

The distribution in Fig.~\ref{uvdata} is
obviously not uniform, showing some concentrations that have been
associated with the Hyades, Pleiades, Sirius and other young moving
groups (see e.g.~Figueras et al.~1997, Chereul et
al.~1997).
A closer examination, however,
reveals an additional pattern of inhomogeneities on a somewhat larger
scale. At least three long {\it branches} 
(we shall call them: 
the Sirius branch, the middle branch and the Pleiades branch, 
respectively from top
to bottom)
can be identified by eye, slightly curved
but almost parallel and running diagonally
across the diagram with a negative slope. We have traced the branches
in a preliminary way by 
following the local maxima (ridge lines) in the $UV$-distribution, as
shown by the dashed lines in Fig.~\ref{uvdata}.

It has been pointed out before (Skuljan et
al.~1997) that the parallax uncertainty can produce
some radially-elongated 
features in the $UV$-plane, since the stars tend to move
radially relative to the zero point ($U=0$, $V=0$) 
when their parallax is changed.
However, this
effect would create some branches
converging to the zero point, which is clearly not the
case in Fig.~\ref{uvdata}. 
We conclude that the parallax uncertainty
cannot be responsible for the distribution found.

Besides the three branches, there seems to exist a fairly sharp `edge line'
at an angle of about $+30^\circ$ relative to the $U$-axis,
connecting the lower-$U$
extremities of the branches 
and defining a region above the line where
only a few stars can be found. Practically all the stars seem to occupy
the lower part of the $UV$-plane bounded by the `edge line' and the
Sirius branch. In such a
situation,
the traditional
velocity-ellipsoid approach does not seem to be appropriate any more.

It would be interesting to compare our distribution from
Fig~\ref{uvdata} to similar diagrams obtained by other authors (Asiain
et al.~1999, Chereul et al.~1998,1999, Dehnen 1998). 
In particular, Fig.~3 of Dehnen
1998 demonstrates all basic features that we introduce here, although
their distribution was obtained {\it without} radial velocities. This clearly 
suggests that the kinematic bias (Binney et al.~1997) does not affect
significantly the inner parts of the $UV$-distribution.

\section{The wavelet transform}

In order to determine the precise nature and the statistical
significance
of the features seen in Fig.~\ref{uvdata}
we have chosen the wavelet transform technique to analyse our data.
There are many examples in the literature
demonstrating how this powerful tool can be used in different areas
(e.g.~Slezak et al.~1990, Chereul et al.~1997), 
but nevertheless 
we shall 
point out some of the basic properties relevant to our work,
concentrating on the two-dimensional case only.

To do a wavelet transform of a function $f(x,y)$ we define a 
so-called {\it analysing wavelet} $\psi(\frac{x}{a},\frac{y}{a})$, 
which is another function (or
another family of functions),
where $a$ is the {\it
  scale} parameter. 
By fixing the scale parameter we can select a wavelet of
a given particular size out of 
a family characterized by the same shape $\psi$.
The wavelet transform $w(x,y)$ is then defined as a {\it correlation} 
function,
so that at any given point ($\xi,\eta$) in the $XY$-plane we have 
one real value for the transform:
\[
 w(\xi,\eta) = 
    \int\limits_{-\infty}^\infty\int\limits_{-\infty}^\infty 
    f(x,y)\,\psi(\frac{x-\xi}{a},\frac{y-\eta}{a})\,dx\,dy,
\]
which is called the {\it wavelet coefficient} at ($\xi,\eta$). Since we
usually work in a discrete case, having a certain finite number of bins
in our $XY$-plane, this means that we shall have a finite number of
wavelet coefficients, one value per bin.

\begin{figure}
   \epsfxsize=\hsize\epsfbox{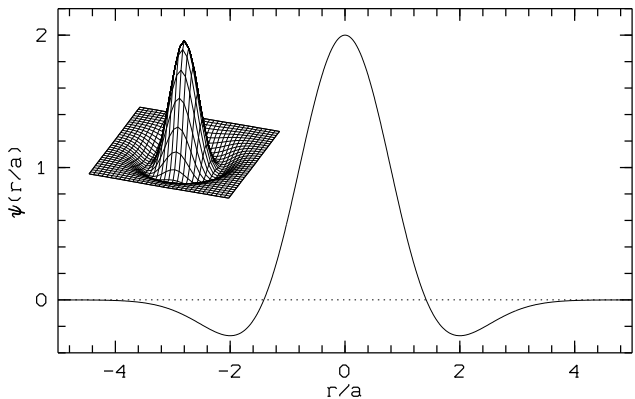}
   \caption{The Mexican hat in two dimensions.}
   \label{radhat}
\end{figure}

The actual choice of the analysing wavelet $\psi$ 
depends on the particular application. When a given data distribution is 
searched for certain {\it groupings} (over-densities) then a so-called {\it
  Mexican hat} is most commonly used (e.g.~Slezak et
al.~1990). A two-dimensional Mexican hat 
(Fig.~\ref{radhat}) is given by:
\[
   \psi(r/a) = \left(2-\frac{r^2}{a^2}\right)e^{-{r^2}/{2a^2}}
\]
where $r^2=x^2+y^2$.
The main property of the function $\psi$ is that the total
volume is equal to zero,
which is what 
enables us to detect any over-densities in our data distribution.
The wavelet coefficients will be all zero if the analysed
distribution
is uniform. But if there is any significant
`bump' in the distribution, the wavelet transform will give a positive
value at that point. Moreover, if we normalize the Mexican hat using a
factor $a^{-2}$, then we will be able to estimate the half-width of the 
`bump', by simply varying the scale parameter $a$: the wavelet
coefficient in the centre of the bump will reach its {\it maximum value} if
the scale $a$ is exactly equal to $\sigma$, assuming that the `bump' is
a gaussian of a form: $\exp(-\rho^2/2\sigma^2)$, $\rho$ being the distance 
from the centre.

Many authors choose the scale 
in such a way that it gives the maximum wavelet
transform. This is an attractive option providing straightforward
information on the
average half-width of the gaussian components in our
distribution. However, there are some situations when we would prefer
somewhat smaller scales, in order to separate two close
components or to detect some narrow but elongated features.

\begin{figure}
   \epsfxsize=\hsize\epsfbox{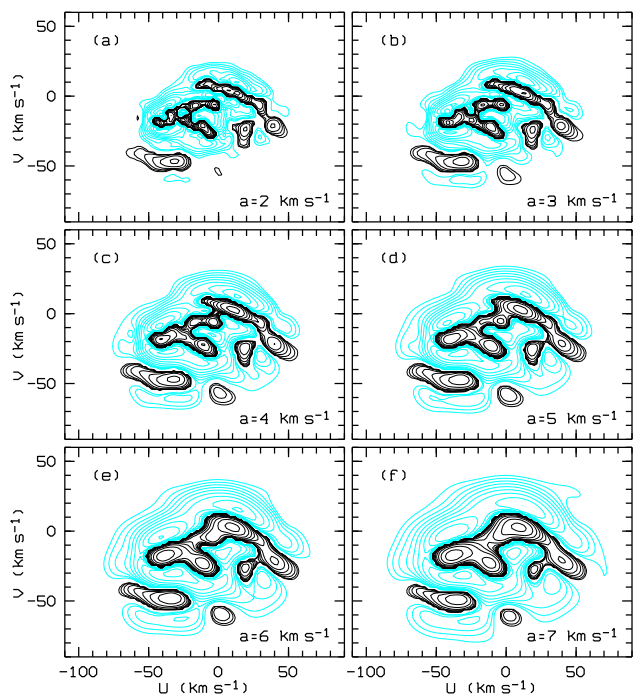}
   \caption{Wavelet transforms of the smooth $UV$-distribution at several
     different scales. Solid lines represent positive contours, while
     dashed lines are used for negative values.} 
   \label{wavdata}
\end{figure}

We have applied several different scales to our $UV$-distribution from
Fig.~\ref{uvdata}b, and the results are shown in Fig.~\ref{wavdata}. 
The positive contours (solid lines) describe the regions where we have an
over-density of stars (grouping), while the negative contours (dashed
lines) show the regions with star deficiency (under-density). We shall
concentrate our attention here only on the positive values. At
about $a\sim4-5$~$\kms$ the normalized wavelet coefficients reach their
maxima around the most populated parts of the $UV$-plane (except the
Hyades moving group, where the maximum wavelet transform occurs somewhere
below $2$~$\kms$, i.e.~at a scale less than the bin size).

\section{Confidence levels}

An important question at this stage is {\it how probable} are the features
revealed in Fig.~\ref{wavdata}, i.e.~what are the
confidence levels for the contours to be above the random noise.
A commonly used procedure to estimate the probabilities is numerical
simulation (sometimes called the Monte 
Carlo method) based on random number generators (see e.g. Escalera \&
Mazure 1992). 

Let us consider again the
distribution in Fig.~\ref{uvdata}b.
By smoothing the probability density function we have found the average
number of stars $\bar N_\ast$ in each bin (to be more precise, we have found 
an optimal estimate of the average value, as close as possible
using the data set available). On the other hand, 
the {\it observed} number of
stars $N_\ast$ in each bin has a {\it statistical uncertainty}. 
This is not only due to
the measurement errors. Even if the velocities have
no random errors (or 
extremely small errors, i.e.~much less than the
bin size) we still have statistical
fluctuations related to the finite sample. 
We can expect that the observed number of stars $N_\ast$ in 
each bin will fluctuate following the Poisson distribution, with an
average of $\bar N_\ast$.
This means that we can regard our observed histogram
as {\it one outcome} from an infinite set of
possibilities, when we let the star counts in every bin fluctuate
according to the Poisson statistics.
We can numerically simulate those ``other possibilities'' and create the
distributions (copies) that ``could have happened''.
If any feature of the distribution is found repeating from one copy to 
another, we can be confident that the feature is real, i.e.~not
generated by noise. Actually, the number of successful appearances
divided by the total number of simulations will give us the probability
of the feature being real.

We use this idea to find the probabilities that our wavelet coefficients 
are {\it positive}, since a positive coefficient is automatically an 
indicator that a grouping 
in the $UV$-plane exists. 
We generate a large number ($N=1000$) of Poisson random copies of our
smooth histogram from Fig.~\ref{uvdata}b. Then, for each of these
copies, we derive a replica of
the original data set, by creating $N_\ast$ stars randomly distributed
over each bin (in total, we shall have a number of stars close to
$n=4597$ for all bins together). 
Finally, we treat the new data set in the same way as the
original: we estimate the density function by applying the
adaptive kernel method\footnote{In order to reduce the computing time,
  we use the original smooth distribution to compute $\lambda_i$
  directly for
  every random data set, and we assume the same optimal smoothing factor.},
and then 
compute the wavelet transform of the corresponding smooth histogram.
This enables us to examine each coefficient ($w$) 
over the whole set of simulations,
and compute the probability
such that the value is positive. If there are 
$N_{\rm p}$ simulations with $w>0$ then we have a probability
$P(w>0) = N_{\rm p}/N$.
This
procedure has been repeated for all wavelet transforms that we shall
present in this paper. 
Typically,  the features shown have a 90 per cent or better
probability of being real.

\section{The data analysis}
\label{analysis}

Although we have started our analysis by examining the whole sample of
stars, we are aware of the fact that the stellar kinematic
properties may depend on the age. It is reasonable to assume that
younger stars have better chances of still keeping the memory of their
original velocities that they acquired at formation. If there is any
grouping in velocity space, and if the grouping is only a result of
cluster evaporation (Eggen's hypothesis), then we would expect to see it
most prominently amongst the youngest stars.

In this paper we are not dealing explicitly with the stellar ages but
we are using the spectral type (colour index actually) 
as an age indicator.
We have divided the whole sample of stars
into two
groups:
\begin{enumerate}
   \item 1036 early-type stars ($B-V<0.3$), and
   \item 3561 late-type stars ($B-V>0.3$).
\end{enumerate}
by choosing $B-V=0.3$ as an arbitrary division point, 
corresponding approximately to the boundary between the A and F
spectral classes. The terms `early-type' and `late-type' should be
regarded here as suitable names to be used in this paper only. Note that 
our late-type group contains spectral classes F and later, with two
distinct clumps (main sequence F stars, plus K giants) as seen in
Fig.~\ref{hrdata}. Analysing the Hipparcos catalogue,
Dehnen \& Binney (1998) found Parenago's
discontinuity at $B-V=0.61$, which means that most of the stars
blue-ward of that are {\it younger} than the Galactic disk
itself. This applies to the F stars in our sample. 
We can conclude that our early-type group (spectral classes
B--F) contains predominantly young stars, while our late-type group 
(F--M) is a mixture of older and younger main-sequence stars, rather
young red-clump stars, and a few old red giants.
The first group should better show the young moving groups and it will
allow us to compare the results with other authors. On the other hand,
the second group should possibly show the old-disk moving groups that can be
compared with Eggen's results.

\subsection{Early-type stars}

\begin{figure}
   \epsfxsize=\hsize\epsfbox{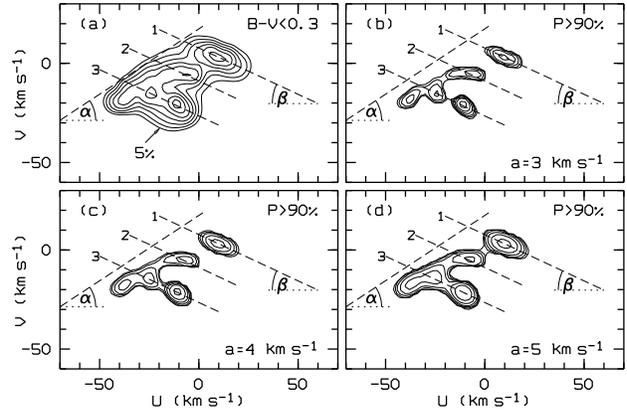}
   \caption{The velocity distribution of early-type stars (a) 
     together with the
     corresponding wavelet transforms (90 per cent confidence) 
     at several different scales (b--d). The dashed lines are used to
     mark the branches and the edge line (see text for equations of lines).}
   \label{uvearly}
\end{figure}

We have applied the adaptive kernel method again to derive the probability
density function $f(U,V)$ for the 1036 early-type stars. An optimal value 
for the smoothing parameter in this case is $h=8.1$~$\kms$.
The smooth distribution,
together with the
corresponding wavelet transforms at several different scales, 
are shown in Fig~\ref{uvearly}.
The three branches are well separated and easily detected. 
Although the
`middle branch' does not appear 
as a feature sufficiently long for a separate 
analysis, we shall nevertheless treat it as an `incomplete' branch.
The maximum wavelet transform is at about $a=4\ \kms$, which also
corresponds to an average half-width of the branches.

\begin{figure}
   \epsfxsize=\hsize\epsfbox{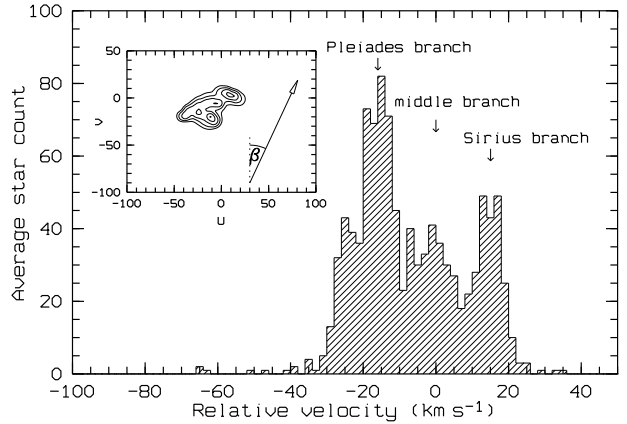}
   \caption{The one-dimensional velocity distribution of early-type
     stars
     along an axis
     perpendicular to the branches (the direction indicated
     in the inserted image).}
   \label{earlycut}
\end{figure}

In order to determine the {\it edge line} we have used the 5-per-cent level
(relative to the maximum) of the
smoothed $UV$-histogram, as indicated in Fig.~\ref{uvearly}a. By fitting a 
straight line to the corresponding portion of the contour we find the
tilt to be $\alpha\approx32^\circ$ ($\tan\alpha\approx0.62$) 
and the edge line to follow the
equation $V=16.4\ \kms+0.62U$.

The tilt of the branches (assuming that they all have the same tilt)
has been computed by rotating the wavelet transform (Fig.~\ref{uvearly}b)
counter-clockwise and examining the distribution along $V$ only
(taking a sum of all positive wavelet coefficients at a given $V$).
At an angle of $\beta\approx25^\circ$ ($\tan\beta\approx0.47$)
the three branches appear as the
narrowest (and strongest)
gaussians, which means that the tilt of the branches in the
$UV$-plane is $-25^\circ$. 
We find the following three equations
for the branches, which are shown in Fig~\ref{uvearly}:
\begin{eqnarray*}
   V_1  = & 7.6\ \kms-0.47U & \mbox{(Sirius branch)}\\
   V_2  = & -8.9\ \kms-0.47U & \mbox{(middle branch)}\\
   V_3  = & -26.6\ \kms-0.47U & \mbox{(Pleiades branch)}
\end{eqnarray*}

It should be noted, however, that these three linear relations describe
the branches well enough only relatively far from the edge line. The branches
seem to curve and follow the edge line at their lower-$U$
extremity. This is especially the case with the Pleiades branch.

In Fig.~\ref{earlycut} we present the one-dimensional distribution
in the direction perpendicular to the branches
($\beta\approx25^\circ$) 
so that each branch appears as a single peak at a fixed
position. 
The zero point of the relative velocity scale has been
centred on the middle branch.
The three branches are approximately equidistant, with a separation of about 
$15\ \kms$.

\subsection{Late-type stars}

\begin{figure}
   \epsfxsize=\hsize\epsfbox{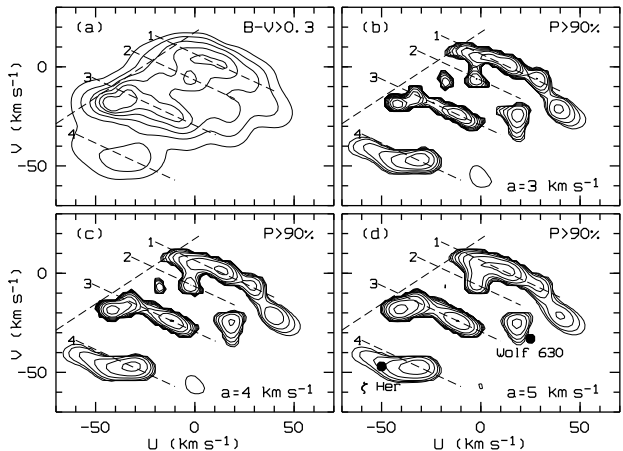}
   \caption{The velocity distribution of late-type stars (a) 
     and the
     corresponding wavelet transforms (90 per cent confidence) 
     at several different scales (b--d). The dashed lines are used to
     mark the branches and the edge line (see text for equations). Two
     Eggen's old-disk moving groups (d) are presented for reference.}
   \label{uvlate}
\end{figure}

The distribution function in the $UV$-plane for the 3561 late-type stars 
has been computed using an optimal smoothing factor of $h=13.6$~$\kms$.
The contour diagrams of the distribution and the
corresponding wavelet transforms at several different scales, 
are shown in Fig~\ref{uvlate}. If we compare this with the early-type
case in Fig.~\ref{uvearly}, we find a similar pattern, although
somewhat more complex.
Besides the three main branches, we have now got some new details, such
as a concentration of stars at about $(20,-30)$~$\kms$ (possibly another
fragment of the middle branch), as well as a new branch in the bottom
part of the diagram, at $(-30,-50)$~$\kms$. In order to get the new
features named, we introduce here two of Eggen's old-disk moving groups,
Wolf~630 and $\zeta$~Herculis, also marked in
Fig.~\ref{uvlate}d. We shall simply use the fact that these two Eggen's moving
groups (Eggen 1965,1971) 
agree well with the features revealed by our wavelet transforms,
although the question of the significance of such a correlation is
still to be answered.

\begin{figure}
   \epsfxsize=\hsize\epsfbox{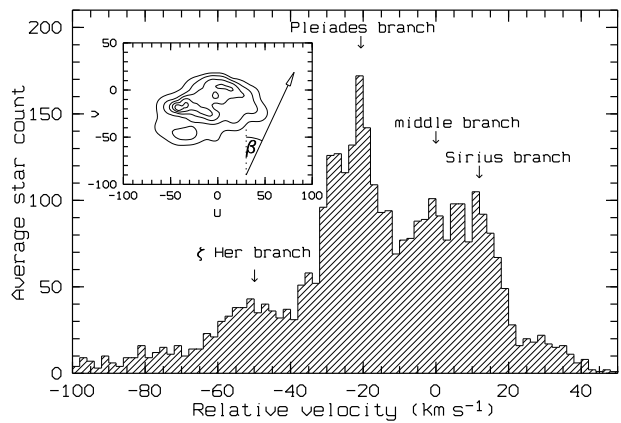}
   \caption{The one-dimensional velocity distribution of late-type
     stars
     along an axis
     perpendicular to the branches (the direction indicated
     in the inserted image).}
   \label{latecut}
\end{figure}

In order to find the positions of the branches in Fig.~\ref{uvlate}, we
could perhaps proceed as when dealing with the early-type stars.
However, the branches now 
seem longer, and possibly curved
(especially the Sirius
branch), so that our procedure for determining the angle $\beta$ does
not seem to be appropriate any more. 
On the other hand, there is not
enough data to do a more sophisticated analysis including the curvature
of the detected features. Our approach was to adopt the same inclination 
angle of $\beta\approx25^\circ$, as derived from the early-type stars, and 
then simply find the positions of the branches from the rotated
one-dimensional
distribution shown in Fig.~\ref{latecut}.
The equations for the branches are:
\begin{eqnarray*}
   V_1  = & 6.9\ \kms-0.47U & \mbox{(Sirius branch)}\\
   V_2  = & -7.0\ \kms-0.47U & \mbox{(middle branch)}\\
   V_3  = & -29.2\ \kms-0.47U & \mbox{(Pleiades branch)}\\
   V_4  = & -62.0\ \kms-0.47U & \mbox{($\zeta$~Herculis branch)}
\end{eqnarray*}

There is also a possible hint of a weak fifth branch
(Fig.~\ref{latecut}) at a relative velocity of about 30~$\kms$.
An overall impression is that the branches are roughly equidistant,
with a separation slightly larger than in the early-type stars. With one 
additional branch,
an average separation is now about 20~$\kms$.
If this `periodicity' is
real, then a two-dimensional Fourier transform of the
distribution will show some peaks in the power spectrum, as we shall
demonstrate in the following section. 

\section{The Fourier transform}

\begin{figure}
   \epsfxsize=\hsize\epsfbox{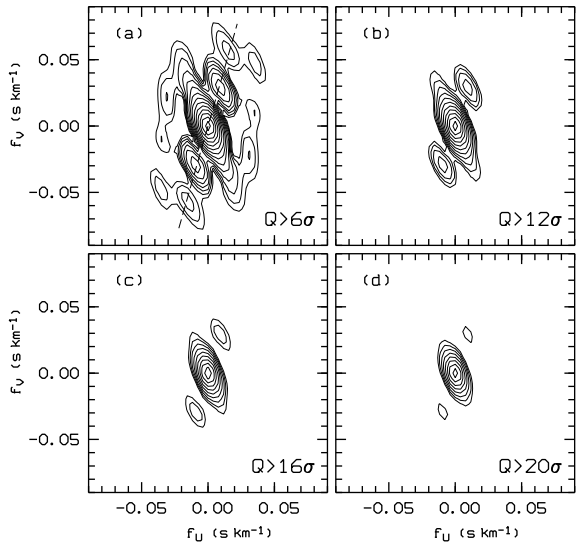}
   \caption{The two-dimensional power spectrum $Q(f_U,f_V)$ 
     of the velocity distribution
     for the whole sample of stars, at several levels of
     significance. The strongest peaks are aligned at
     $\gamma\approx74^\circ$, as indicated by the dashed line (a).}
   \label{fftdata}
\end{figure}

The two-dimensional power spectrum $Q(f_U,f_V)$
of the smooth $UV$-histogram for 
the whole sample of stars
(square root of the power spectral density) 
is shown in Fig.~\ref{fftdata}.
Most of the total power is concentrated within the central 
bulge (maximum power density of about 4700), 
corresponding to a roughly gaussian distribution of the stars in 
the $UV$-plane. There are also two relatively strong side peaks 
(maximum power density of about 920)
symmetrically arranged 
around the central bulge at frequencies ($\pm0.008,\pm0.031$)~$\skm$,
as well as two higher harmonics at ($\pm0.016,\pm0.054$)~$\skm$.
The  peaks are arranged along 
a straight line at an
angle of $\gamma=74^\circ$
(the dashed line in Fig.~\ref{fftdata}a). Some other features can also
be seen at relatively high significance levels, but we shall concentrate 
here only on the aligned peaks.
They define a planar
wave in the velocity plane.
Of course, the peaks at negative frequencies 
are simply
symmetrical images of the positive ones, without any
additional information.

In order to estimate the significance of the features in the power
spectrum, we have proceeded in a similar way as when treating the wavelet
transforms. A large number of random copies have been used to see how
the power spectral density fluctuates at any given frequency point. The
standard deviation $\sigma$, has been computed for each bin and
the ratio $Q/\sigma$ has been used as a measure of significance. 

Every peak in the power spectrum can be related to a planar wave
propagating in a certain direction, with a frequency $f=\sqrt{f_U^2+f_V^2}$.
We find a period of about 33~$\kms$ for the stronger side peak (the one
closer to the central bulge), and about 17~$\kms$ for the first
harmonic. These values are in a good agreement with our wavelet transform
analysis:
the longer period corresponds to the separation between the two
most prominent branches in the $UV$-distribution (Sirius and Pleiades),
while the second one can be related to the remaining weaker branches.
It should be noted, however, that the power spectrum contains 
obviously much more information
than we have extracted here, and a more detailed analysis is needed.

\section{Conclusion}

Using the Hipparcos astrometry and published radial velocities,
we have undertaken
a detailed examination of the $UV$-distribution of stars in the solar
neighbourhood. This analysis reveals a branch-like
structure both in early-type and late-type stars, 
with several branches running diagonally with a negative slope relative to
the $U$-axis. The branches are seen at relatively high
significance levels (90 per cent and higher) when analysed using the wavelet
transform technique. They are roughly equidistant in velocity space,
as confirmed by 
the two-dimensional power spectrum.

The branch-like velocity distribution may be due to
the galactic spiral structure itself, or some other global 
characteristics of the galactic potential
combined with the initial velocities at the time of
star formation. A possibility also exists that this is a result of a sudden
burst of star formation that took place some time ago in several
adjacent spiral arms. What we see now in the velocity space might be an
image of the galactic spiral arms from real space. 
The main problem with
this hypothesis, however, is that the stars in our $UV$-branches are not of
the same age. Some groups (like the Pleiades)
are even composed of stars having a range of
different ages (Eggen 1992, Asiain et al.~1999).

There are other possibilities that we are currently testing by means of
numerical simulations involving the motion of stars in the galactic
potential (including the spiral component). Some of the details in the
velocity-plane structure can be simulated by choosing appropriate
initial velocities for stars being created in the galactic spiral arms,
in combination with some velocity dispersion.
We are going to elaborate these ideas in more detail in a future paper.

\section*{Acknowledgments}

This work has been supported by the University of Canterbury Doctoral
Scholarship and by the Royal Society of New Zealand 1996 R.H.T.~Bates
Postgraduate Scholarship to JS, as well as by a Marsden Fund grant to
the Astronomy Research Group at the University of Canterbury.

We gratefully acknowledge the comments of an anonymous referee who
directed us towards the use of the adaptive kernel method.

It was with great sadness that we heard of the passing of Olin Eggen
during the final stages of this work. His insight into these problems
was an inspiration to the field.

\label{lastpage} 

\end{document}